\documentclass[aps,prb,twocolumn,floats,showpacs,superscriptaddress]{revtex4-1}
\usepackage{graphicx,epsfig}
\usepackage{times}
\usepackage{graphics,dcolumn,bm,float}
\usepackage{amssymb,amsmath,rotate,color}
\usepackage[title,titletoc,toc]{appendix}

\usepackage[pagebackref=false,colorlinks,linkcolor=blue,citecolor=blue,urlcolor=magenta]{hyperref}

\begin{document}
\unitlength 1 cm
\newcommand{\be}{\begin{equation}}
\newcommand{\ee}{\end{equation}}
\newcommand{\bea}{\begin{eqnarray}}
\newcommand{\eea}{\end{eqnarray}}
\newcommand{\nn}{\nonumber}
\newcommand{\vk}{\vec k}
\newcommand{\vp}{\vec p}
\newcommand{\vq}{\vec q}
\newcommand{\vkp}{\vec {k'}}
\newcommand{\vpp}{\vec {p'}}
\newcommand{\vqp}{\vec {q'}}
\newcommand{\bk}{{\bf k}}
\newcommand{\bp}{{\bf p}}
\newcommand{\bq}{{\bf q}}
\newcommand{\br}{{\bf r}}
\newcommand{\bR}{{\bf R}}
\newcommand{\up}{\uparrow}
\newcommand{\down}{\downarrow}
\newcommand{\cdag}{c^{\dagger}}
\newcommand{\hlt}[1]{\textcolor{red}{#1}}

\title{Strong coupling approach to Mott transition of massless and massive Dirac fermions on honeycomb lattice}
\author{E. Adibi}
\address{Department of Physics, Sharif University of Technology, Tehran 11155-9161, Iran}
\author{S. A. Jafari}
\thanks{\href{mailto:akbar.jafari@gmail.com}{akbar.jafari@gmail.com}}
\address{Department of Physics, Sharif University of Technology, Tehran 11155-9161, Iran}
\address{Center of excellence for Complex Systems and Condensed Matter (CSCM), Sharif University of Technology, Tehran 1458889694, Iran}
\affiliation{School of Physics, Institute for Research in Fundamental Sciences, Tehran 19395-5531, Iran}

\begin{abstract}
 Phase transitions in the Hubbard model and ionic Hubbard model at half-filling on the honeycomb lattice are investigated in the strong coupling perturbation theory which corresponds to an expansion in powers of the hopping $t$ around the atomic limit. 
Within this formulation we find analytic expressions for the single-particle spectrum, whereby the calculation of 
the insulating gap is reduced to a simple root finding problem. This enables high precision determination of the insulating gap
that does not require any extrapolation procedure.
The critical value of Mott transition on the honeycomb lattice is obtained to be $U_c\approx 2.38 t$. Studying the ionic Hubbard model at the lowest order, we find two insulating states, one with Mott character at large $U$ and 
another with single-particle gap character at large ionic potential, $\Delta$. The present approach gives a 
critical gapless state at $U=2\Delta$ at lowest order. By systematically improving on the perturbation expansion,
the density of states around this critical gapless phase reduces. 

\end{abstract}
\maketitle
\section{Introduction}
Graphene is the most extensively studied -- both theoretically and experimentally -- example
of a Dirac solid where the effective motion of charge carriers is described by the Dirac theory
in 2 spatial dimensions. The Dirac theory in graphene is a continuum limit of a simple tight-binding
hopping Hamiltonian on a honeycomb lattice of graphene material. Breaking the 
sublattice symmetry of the underlying honeycomb lattice leads to a mass in the Dirac theory. 
Recent engineering of the band gap in graphene on SiC brings the study of both massless
and massive Dirac fermions into the frontier of graphene research~\cite{Conrad}.
Therefore graphene is a natural framework to study both massive and massless Dirac fermions in 2+1 
(space+time) dimensions. The atom next to Carbon in the column IV of the periodic table is Si
which also has a two-dimensional allotrope known as silicene with a honeycomb structure, albeit
with a larger lattice constant than graphene. Corresponding to larger distance, the hopping amplitude
between the neighboring atoms in silicene will be smaller than graphene. 
Already for the case of graphene, the ab-initio estimates of the Hubbard $U$ gives a value 
near $10$ eV, making a ratio of $U/t\sim 3.3$~\cite{Wehling}. This ratio is even larger in silicene
due to larger lattice constant and hence a smaller hopping amplitude. In the case of silicene
the ratio is given by $U/t\approx 4.2$~\cite{KatsnelsonSilicene}. Given such large values
of $U/t$ in both graphene and silicene where the low-energy effective theory is a Dirac theory,
it is necessary to understand the effect of such large values of Hubbard $U$ on the electronic
properties of two dimensional Dirac fermions.

A natural framework to approach from the infinite $U$ side is strong coupling perturbation 
theory to expand in powers of $t/U$. This can be done in two ways: (1) is to do brute force
perturbation theory~\cite{Metzner}, and the other way is to use a dual transformation and to
rewrite the strongly correlated Hamiltonian in terms of dual degrees of freedom~\cite{Senechal98}.
We find the later approach rewarding as it clearly indicates the onset of gap closing by approaching
from the strong coupling side. Despite some pathologies in the analytic continuation, in the lower
orders of perturbation theory considered here, we are able to obtain closed form formulae for the
spectral functions without encountering the problems of analytic continuation faced by earlier
investigators. Within this approach we identify the Mott transition in the half-filled
massless Dirac sea at zero temperature. Given our analytic formulae for the spectral functions,
the determination of Mott gap is reduced to a simple root finding problem that can be 
done with arbitrary precision. This does not require extrapolation procedure~\cite{Hassan}.
Approaching from the Mott side one might think that
electrons being localized in the Mott phase, do not have any idea what is going to happen
when the Hubbard $U$ is reduced. However on the weak coupling side we know that the underlying 
honeycomb structure leads to Dirac spectrum. Therefore the question would be how does the 
Mott phase know that upon reducing the Hubbard $U$ it should become a Dirac solid? Interesting
picture that emerges within the present dual transformation approach is that deep in the Mott
phase, the dual fermions have a Dirac cone structure, albeit far-away from the Fermi level
within the high energy states of upper and lower Hubbard bands, and hence the Dirac "genome" is passed
across the quantum critical point separating the Dirac liquid and the Mott insulating phase. 

We also take the same approach to study the massive Dirac fermions approaching from the Mott side. 
For this model, the $U/t$ is not the only parameter governing the phase diagram. The presence of
another energy scale $\Delta$ related to gap (mass) makes it more complicated. 
In the large $U$ limit again we have the Mott phase. When the Hubbard $U$ is negligible 
in comparison to $\Delta$, its main effect is to renormalize Fermi liquid parameters of the
underlying metallic state, and hence the relevant parameter $\Delta$ opens up a single-particle
gap and we have a band insulator~\cite{HafezJafari}. For the intermediate regime our
earlier dynamical mean field theory (DMFT) study suggests the presence of a gapless semimetallic
state which is born out of the competition between the two parameters $U$ and $\Delta$~\cite{HafezPRB,Ebrahimkhas}. 
Within the present approach at the lowest orders of the kinetic energy $t$, we find that
there is a critical point separating the Mott and band insulating phases. The system
at this critical point is gapless and corresponds to a semimetallic (Dirac cone) state.
Systematically improving the perturbation theory by going to higher orders shows that the
density of states (DOS) around this quantum critical semimetallic states tends to deplete.

The paper is organized as follows. We begin by reviewing the strong coupling  expansion method in section \ref{method}. In section \ref{HModel} the method is applied to the half-filled Hubbard model and by using an analytic approach the  critical interaction of the Mott transition is obtained. The method is also employed to investigate the possible phases of the half-filled ionic Hubbard model in section \ref{IH-model}. 
Finally our findings are summarized and conclusions are drawn in section~\ref{discussion}. The paper is accompanied by two appendices which present the expression for DOS on honeycomb lattice and our formulae for self-energies of auxiliary fermions in the ionic Hubbard model case.

\section{Method Of Calculation \label{method}}
We employ the strong coupling expansion to study the semimetal to insulator transition (SMIT) of 
the Hubbard model on the honeycomb lattice. We also use the method to characterize phase diagram 
of the ionic Hubbard model at zero temperature. In what follows, we briefly describe the 
method proposed in Ref.~\onlinecite{Senechal98}. Generally speaking, in the strong coupling limit, the 
Hamiltonian is written as the sum of the unperturbed local Hamiltonian $H_0$ and the perturbation $H_1$:
\be
H=H_0+H_1.
\ee
According to formulation of Ref.~\onlinecite{Senechal98}, $H_0=\sum_{i}\ h_i(\cdag_{i\sigma},c_{i\sigma})$ 
where $H_0$ is diagonal in variable $i$ and $\sigma$ denotes all the other variables of the problem. 
If we assume $i$ as site variable, $H_0$ is written as a sum over on-site Hamiltonians $h_i$. 
On the other hand $H_1$ is supposed to be a one-body hopping operator 
$H_1=\sum_{ij}\sum_{\sigma}\ V_{ij}\cdag_{i\sigma}c_{j\sigma}$ where the Hermitian matrix
$V$ is the hopping amplitude between orbitals located at sites $i,j$.
The partition function in the path-integral formulation can then be expressed as,
\bea
Z\!\!&=&\!\!\int [d\gamma^\star\ d\gamma]\ \exp\bigg[ -\int_{0}^{\beta}d\tau\bigg\lbrace \sum_{i\sigma} \gamma^{\star}_{i\sigma}(\tau)(\partial_\tau-\mu)\gamma_{i\sigma}(\tau)\nn\\
&+&\sum_{i}h_i(\gamma^{\star}_{i\sigma}(\tau),\gamma_{i\sigma}(\tau))+\sum_{ij\sigma}\gamma^{\star}_{i\sigma}(\tau)V_{ij}\gamma_{j\sigma}(\tau) \bigg\rbrace\bigg],
\eea
where $\gamma_{i\sigma}(\tau)$, $\gamma^{\star}_{i\sigma}(\tau)$ denote to imaginary Grassmann fields 
of the electrons and $\beta$ is inverse of temperature $T$.
In the Hubbard-like models $H_0$ is not quadratic, hence the simple form of an ordinary Wick theorem 
can not be used to construct a diagrammatic expansion for the Green's functions\footnote{Although a
more complicated version of the Wick theorem still exists. But it is not easy or intuitive to work
with}. Introducing the auxiliary Grassmann fields 
$\psi_{i\sigma}(\tau)$, $\psi^\star_{i\sigma}(\tau)$, via the Grassmann version of the 
Hubbard-Stratonovich transformation~\cite{HS transformation} we can write,
\bea
&&\int [d\psi^\star d\psi] \exp\bigg[\int_{0}^{\beta} d\tau \sum_{i\sigma} \bigg\lbrace\sum_j \psi^\star_{i\sigma}(\tau) (V^{-1})_{ij} \psi_{j\sigma}(\tau)\nn\\
&&\qquad\qquad+\psi^\star_{i\sigma}(\tau)\gamma_{i\sigma}(\tau)+\gamma^\star_{i\sigma}(\tau)\psi_{i\sigma}(\tau)\bigg\rbrace\bigg]\nn\\
&&=\det(V^{-1})\exp\bigg[-\int_{0}^{\beta} d\tau \sum_{ij\sigma} \gamma^\star_{i\sigma}(\tau) V_{ij} \gamma_{j\sigma}(\tau) \bigg].
\eea
With the aid of this equation the the partition function can be rewritten as,
\be
Z=\int[d\psi^\star d\psi] \exp\bigg[-\bigg\lbrace S_0[\psi^\star,\psi]+\sum_{R=1}^{\infty} S^R_{int}[\psi^\star,\psi]\bigg\rbrace \bigg],
\ee
where the action has a free auxiliary fermion part given by the inverse of the hopping matrix of 
original fermions, 
\bea
S_0[\psi^\star,\psi]=-\int_{0}^{\beta}\ d\tau \sum_{ij\sigma} \psi^\star_{i\sigma}(\tau)\ (V^{-1})_{ij}\ \psi_{j\sigma}(\tau),
\eea
and an infinite number of interaction terms
\bea
 &&S^{R}_{int}[\psi^\star,\psi]=\frac{-1}{(R!)^2}\sum_{i}\sum_{\lbrace\sigma_l\sigma^{\prime}_l\rbrace} \int_{0}^{\beta}\ \prod_{l=1}^{R}\ d\tau_l d\tau'_l\nn\\ &\times &\psi_{i\sigma_1}^{\star}(\tau_1)\ldots\psi_{i\sigma_R}^{\star}(\tau_R)\psi_{i\sigma'_R}(\tau'_R)\ldots\psi_{i\sigma'_1}(\tau'_1)\nn\\
&\times &\bigg\langle \gamma_{i\sigma_1}(\tau_1)\ldots\gamma_{i\sigma_R}(\tau_R)\gamma^{\star}_{i\sigma'_R}(\tau'_R)\ldots\gamma^{\star}_{i\sigma'_1}(\tau'_1)\bigg\rangle_{0,c}.
\eea
The above equation denotes a vertex with $R$ incoming $\psi$ fermions and $R$ outgoing $\psi$ fermions. 
Note again that $\psi$ fermions are auxiliary (dual) fermions. Thinking in terms of $\psi$ fermions,
now their kinetic energy scale is given by $V^{-1}$ which is a large number when the kinetic energy $V$ 
of the original fermions is much smaller than the Coulomb energy scale $U$. 
Therefore standard diagrammatic perturbation theory can be applied. The only (very important) difference
with the text book diagrammatics will be that in the present case the vertex is not a simple number,
but acquires a non-trivial dynamic structure given by the the cumulant average 
$\langle\cdots\rangle_{0,c}$ of the original Grassmann fields. These are the connected correlation 
functions of original fermions with respect to the local Hamiltonian $h_i$. Higher order cumulants
are expected to be less important in the limit of large $U$. Once we have some lower order cumulants
which only depend on the form of the local Hamiltonian $h_i$, the cumulants can be calculated
straightforwardly~\cite{Sherman}. Once the multi-particle cumulants of the original fermions
are known, they act as dynamic vertices for the auxiliary fermions and from this point, 
we can use standard perturbation theory for the auxiliary fields. Eventually, If $G$ denotes the 
Green's function of the original fermions and $\Gamma$ the self-energy of the auxiliary fermions, 
the relation between them is given by~\cite{Senechal2000},
\be
G=(\Gamma^{-1}-V)^{-1}\label{green'sfunction}
\ee
This means to obtain Green's function, we have to compute the self-energy $\Gamma$ 
of the auxiliary fermions which can be done with standard perturbation theory.
Further details of the method are given in Refs. \onlinecite{Senechal98, Senechal2000} and will not be repeated here.

In the following sections we apply the method presented here to two models at half-filling 
on the honeycomb lattice, namely the Hubbard model and ionic Hubbard model. 
On the honeycomb lattice, free propagator of the auxiliary 
fermions is given by the inverse of
\bea
 V(\mathbf{k})=
 \begin{pmatrix}
 0 & ts(\mathbf{k}) \\ ts^{\star}(\mathbf{k}) & 0
 \end{pmatrix}
\eea
where $\mathbf{k}=(k_x,k_y)$, and 
$s(\mathbf{k})=\exp(-\mathrm{i}k_x a)+2\exp(\frac{\mathrm{i}k_x a}{2})\cos(\frac{\sqrt{3}k_y a}{2})$. 
The atomic separation of the honeycomb lattice assumed to be $a=1$. The $2\times 2$ matrix structure
comes from the two sublattice structure of the honeycomb lattice. In the following we use the 
standard pertubation theory to study the Hubbard and ionic Hubbard models the non-interacting
limit of which corresponds to massless and massive Dirac fermions.

\section{Hubbard Model\label{HModel}}
\begin{figure}[tb]
    \includegraphics[width=3cm]{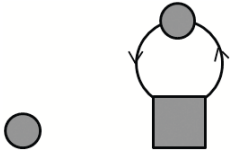}
    \caption{Diagrams contributing to the self-energy of the auxiliary fermions up to order $t^2$.}
    \label{diagram}
\end{figure}

The Hubbard Hamiltonian for spin-1/2 fermions is given by,
\be
H=-t\sum_{\langle ij\rangle,\sigma}(\cdag_{i\sigma}c_{j\sigma}+\mathrm{h.c})+U\sum_i\ n_{i\uparrow}n_{i\downarrow}-\mu\sum_{i\sigma}n_{i\sigma}
\ee
where $\cdag_{i\sigma}$ ($c_{i\sigma}$) creates (annihilates) a fermion of spin 
projection $\sigma=\uparrow,\downarrow$ on lattice site $i$, 
$n_{i\sigma}=\cdag_{i\sigma}c_{i\sigma}$,  $t$ denotes the nearest-neighbor hopping amplitude 
and $U\geq 0$ denotes the strength of the on-site repulsion.  
Through the paper, we focus at half-filling \big($\sum_{\sigma}\langle n_{i\sigma}\rangle=1$\big) 
by setting the chemical potential $\mu$ to $U/2$. For the strong coupling expansion of the
Hubbard model, $H_0$ corresponds to the atomic limit and $H_1$ is equivalent to the kinetic term. 
The diagrams contributing to $\Gamma$ 
up to order $t^2$ are presented in Fig.~\ref{diagram} which lead to the following expression for 
$\Gamma$~\cite{Senechal2000},
\be
\Gamma(i\omega)=\bigg(\frac{i\omega}{(i\omega)^2-(U/2)^2}+\frac{3.45\ t^2 (U/2)^2 (i\omega)}{((i\omega)^2-(U/2)^2)^3}\bigg)\ \mathbb{I}\label{self1}
\ee
where $i\omega$  denotes to a complex frequency  and $\mathbb{I}$ stands for $2\times 2$ identity matrix
in the space of two sublattices.

As is evident from the above self-energy (for more details see Ref.~\onlinecite{Senechal98}), the above
self-energy violates the casuality. A casual Green's function (or self-energy) is Lehmann representable 
if it can be written as a Jacobi continued fraction form. So one has to find out a Jacobi continued 
fraction form of self-energy Eq.~(\ref{self1}) which in this case is simple and turns out to be
\bea
\Gamma(i\omega)=\cfrac{1}{i\omega-\cfrac{(U/2)^2}{i\omega-\cfrac{3.45\ t^2}{i\omega-\cfrac{(U/2)^2}{i\omega}}}}\ \ \mathbb{I}\label{self2}
\eea
which is equivalent to Eq.~\eqref{self1} up to $(t/U)^2$. 
Now we can calculate the Green's function by substituting self-energy into 
Eq.~\eqref{green'sfunction}. In order to monitor the Mott transition, we should calculate the 
DOS $\rho(\omega)=-\frac{1}{\pi}\lim_{\eta\rightarrow 0^{+}} \sum_{\mathbf{k}} \mathrm{Im}\ \mathrm{Tr}\ G(\mathbf{k},\omega+i\eta)$ at different interaction strength. In other words, to 
identify the electronic properties of the system by increasing $U$, we calculate the single-particle 
gap that extracted from DOS by integration over wave vectors numerically. But in doing so,
it is hard to judge when the gap opens by increasing $U$ due to artificial Lorentzian broadening $\eta$
used in the Greens' functions to avoid numerical divergences. As we will explain shortly in the following, we are able to
work out the integration analytically which enables us to avoid both numerical errors as well as 
the continued fraction issues. This reduces the determination of the Mott gap into a simple root finding
problem which can be solved with arbitrary precision.

Assuming the self-energy of auxiliary fermions in Eq.~\eqref{self1} ( or Eq.~\eqref{self2} ) has a more general form 
$\Gamma=\mathcal{F}(i\omega)\mathbb{I}$ and plugging in Eq.~\eqref{green'sfunction}, the 
DOS of physical electrons is given by,
\bea
\rho(\omega)=-\frac{1}{\pi}\lim_{\eta\rightarrow 0^{+}} \sum_{\mathbf{k}} \mathrm{Im}\bigg[\ \frac{1}{1/\mathcal{F}(\omega+i\eta)-t|s(\mathbf{k})|}\nn\\
+\ \frac{1}{1/\mathcal{F}(\omega+i\eta)+t|s(\mathbf{k})|}\ \bigg] \label{interactind-dos}
\eea 
On the other hand, in non-interacting honeycomb lattice (graphene) DOS 
of a hopping Hamiltonian is given by~\cite{hobson},
\bea
\rho_0(\omega)&=&-\frac{1}{\pi}\lim_{\eta\rightarrow 0^{+}} \sum_{\mathbf{k}} \mathrm{Im}\bigg[\ \frac{1}{\omega+i\eta-t|s(\mathbf{k})|}\nn\\
&&\qquad\qquad+\ \frac{1}{\omega+i\eta+t|s(\mathbf{k})|}\ \bigg]\nn\\
&=& \frac{|\omega|}{\pi^2}\ \frac{1}{\sqrt{Z_0}}\ K\big(\sqrt{\frac{Z_1}{Z_0}}\big)\label{graphene-dos}
\eea 
where
\bea
Z_0=\left\lbrace
\begin{array}{cc}
(1+|\omega|)^2-(\omega^2-1)^2/4\qquad\qquad &  |\omega| < 1\\
\\
4|\omega|\qquad & 1 \leq |\omega| \leq 3 
\end{array} \right.\label{z0}
\eea
and
\bea
Z_1=\left\lbrace
\begin{array}{cc}
 4|\omega|\qquad&  |\omega| < 1\\
\\
(1+|\omega|)^2-(\omega^2-1)^2/4\qquad\qquad & 1 \leq |\omega| \leq 3 
\end{array} \right.\label{z1}
\eea
Here $K(x)$ is the complete elliptic integral of first kind \cite{elliptic}. 
This representation is valid as long as the imaginary part of the argument
passed into the above function is negligible. 
Comparison of Eqns.~(\ref{interactind-dos}) and (\ref{graphene-dos}), DOS of interacting
problem is analytically obtained as 
\be
   \rho(\omega)=\rho_0({\cal F}^{-1}(\omega))
   \label{rhoF}
\ee
This representation is valid as long as ${\cal F}^{-1}$ has a 
negligible imaginary part. For the pure Hubbard model it turns out 
that when ${\cal F}^{-1}$ is evaluated at $\omega+i0^+$, its imaginary 
part tends to zero. Therefore the above representation is valid. 
A nice property of the function $\rho_0$ is that it vanishes when its
argument $|\omega|$ exceeds $3$. This statement is exact and involves no numerical 
errors. Therefore for the pure Hubbard model, the gap opening corresponds
to the condition
\be
   |{\cal F}^{-1}(\omega+i0^+)|>3
   \label{gapcond1.eqn}
\ee

Based on particle-hole symmetry, we expect the Mott-Hubbard gap to open up at
$\omega=0$, we only need to monitor the behavior of ${\cal F}^{-1}$ at $\omega=0$. When 
this quantity is larger than $3$, DOS is zero, and hence we have a gap. Therefore
starting from the large $U$ side it only suffices to monitor the function ${\cal F}^{-1}$
at $\omega=0$ for various values of $U$. Upon reducing $U$ once this values drops below $3$
indicates that we have entered the conducting phase. 
Therefore we have reduced the problem of determination of the Mott gap into 
a root finding problem indicated by Eq.~\eqref{gapcond1.eqn} which can be
solved with arbitrary precision at negligible computational cost.
In this work we determine the gaps up to the precision of $10^{-4}t$.
Note that within the methods based on Jacobi continued fraction 
followed by numerical integration scarcely can get such resolutions.

\begin{figure}[t]
    \includegraphics[width=8.3cm]{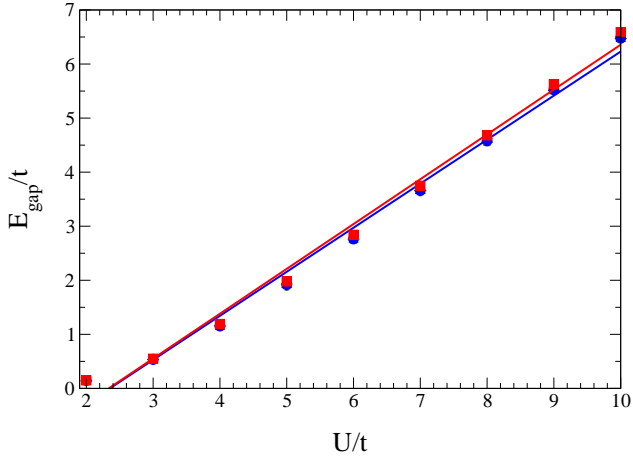}
    \caption{ (Color online)
    $U$ dependence of the single-particle gap calculated from
    Eq.~\eqref{gapcond1.eqn}. For the self-energy ${\cal F}(\omega+i0^+)$ we
    can use both Eq.~(\ref{self1}) (red squares) and Eq.~(\ref{self2}) (blue circles). 
    The red (blue) line indicates the best fitted line to 
    red (blue) data.}
    \label{gap-hubbard}
\end{figure}

Let us now implement condition~\eqref{gapcond1.eqn} to study the Mott transition on the
honeycomb lattice.
In Fig.~\ref{gap-hubbard} the single-particle gap $E_{\rm gap}$ as extracted from 
Eq.~\eqref{gapcond1.eqn} versus on-site interaction $U$ is shown. 
As we study the Mott transition from strong coupling limit, the 
single-particle gap is 
determined for large interaction strengths from the behavior of $\mathcal{F}(\omega)$
i.e. the self-energy of auxiliary fermions. Now to evaluate this self-energy, we have 
two options: One is to use Eq.~(\ref{self1}) and the other is to use the 
continued fraction form Eq.~(\ref{self2}). Note that this options are not available 
in the absence of analytic formula for DOS. 
As can be seen in Fig.~\ref{gap-hubbard},  
the two procedures agree on the value of Hubbard gap obtained from the condition \eqref{gapcond1.eqn}. 
To characterize the critical Coulomb interaction $U_c$ for the SMIT, we do not bother with
extrapolations of the Lorentzian width of the numerical integration
as the limit $\eta\to 0$ has been properly encoded in Eq.~\eqref{graphene-dos}.
We approach from
the Mott side where we are sure that (1) the method is more reliable as it is a perturbation
from the Mott side, and (2) the gap is clearly open. Then having a number of data in the Mott
side, we extrapolate by fitting the data to find out the value of $U$ at which $E_{\rm gap}$
extrapolates to zero. With this approach we find $U_c=2.38t$ for the Mott transition within
the present second order strong coupling approximation. For $U>U_c$, the gap is well fitted 
by linear function of $U$ that is the canonical behavior of a correlation driven Mott insulator
for large $U$. This is not surprising as the method builds in the assumption of large $U$ by
expansions in powers of $t/U$. 

\begin{figure}[t]
    \includegraphics[width=8.3cm]{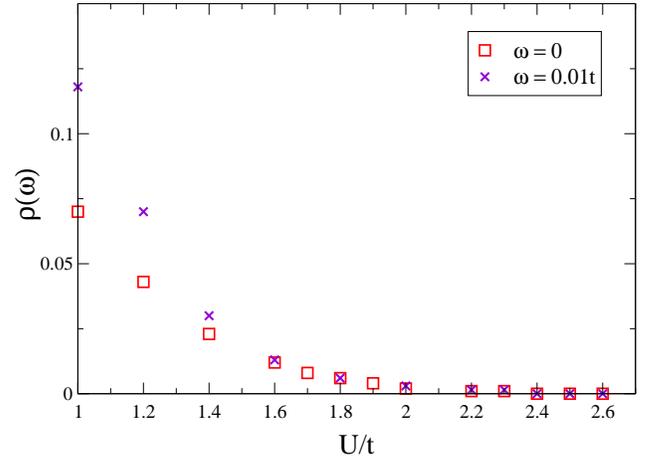}
    \caption{ (Color online)
    The $\eta\to 0$ extrapolated $\rho(\omega)$  for $\omega=0$ and $\omega=0.01 t$ as a function of the on-site 
    Coulomb interaction U on the honeycomb lattice. }
    \label{gap-eta}
\end{figure}

To demonstrate the advantage of the present analytical approach let us see how the previous 
authors find out the onset of gap formation. First for a small but finite value of $\eta$ the
integration required in Eq.~(\ref{interactind-dos}) is calculated by numerical integration over 
wave vectors of the first Brillouin zone of the honeycomb lattice. Thus, one computes 
{$\rho^\eta(\omega)$} for a few values of the Lorentzian broadening parameters $\eta$ at $\omega=0$.
Then by means of polynomial fitting one extrapolates to $\eta\to 0$ limit. The extrapolated 
$\rho(\omega=0)$  must vanish in the insulating phase. 
However this is ambiguous, because even in the Dirac (non-Mott) phase the DOS at $\omega=0$ 
is expected to be zero. To somehow get around this, it was suggested to focus on the DOS at slightly
different energy scale, e.g. $\omega=0.01t$~\cite{Hassan}. 
We have presented a comparison of these two in Fig.~\ref{gap-eta}. This figure suggests 
that the Mott phase is stabilized for $U \geq 2.4 t$. However the non-zero DOS at $\omega=0$
in the semimetallic side is not remedied.

\begin{figure}[t]
  \vspace{0.3cm}
    \includegraphics[width=8cm]{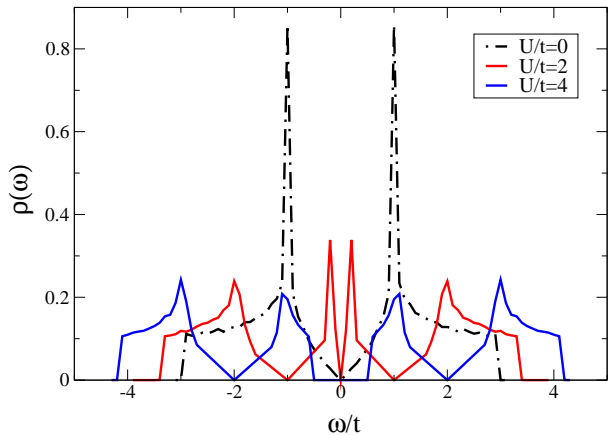}
    \caption{(Color online) DOS of the Hubbard model on honeycomb lattice. 
    DOS at $U=0$ is given for reference.}
    \label{dosinteraction}
\end{figure} 
Now let us employ the present analytical formula to study the profile of the DOS as a function of 
the Hubbard $U$. In Fig.~\ref{dosinteraction} we plot DOS obtained from Eq.~\eqref{rhoF}. 
The non-interacting DOS has been denoted by dotted line for reference. As can be seen in the 
semimetallic phase there is one linear DOS feature at $\omega=0$ which is due to the Dirac cone
at the K points of the Brillouin zone. But in addition there appear another valley in DOS which would 
correspond to Dirac cone at higher energy scales corresponding to $\omega=U/2$. Interestingly
this feature survives in the Mott phases where the major low-energy Dirac cone has been gapped out
by strong $U$. This can be easily understood from the form of Eq.~\eqref{self1}:
As can be seen the auxiliary fermion self-energy $\Gamma$ diverges at $\omega=\pm U/2$ which
corresponds to ${\cal F}^{-1}=0$. But from Eq.~\eqref{rhoF}, when the argument of bare $\rho_0$ 
becomes zero, it will correspond to a Dirac node. 
This feature may help to sheds light on the meaning of auxiliary fermions: The divergence in the self-energy of 
auxiliary fermions corresponds to Dirac nodes of the original electrons. 

At this point let us emphasize that 
{\em the expression of DOS in terms of a function ${\cal F}$ is quite
general}.  This is because the expansion is basically in powers of the hopping matrix
$V(\mathbf{k})$ which is a combination of Pauli matrices. But since odd powers of the
Pauli matrices do not survive the trace, only even powers corresponding to even orders
of perturbation expansion survive the trace needed in calculation of DOS. 
Therefore at any (even) order of perturbation theory, DOS can be expressed
in the form of Eq.~\eqref{rhoF}. Going to higher orders only improves the dynamical
structure of the function $\cal F(\omega)$. 

\section{Ionic Hubbard Model\label{IH-model}}
\begin{figure}[tb]
    \includegraphics[width=5cm]{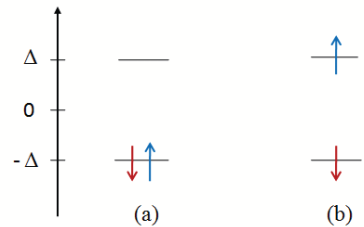}
    \caption{Atomic limit schematic representation of (a) $U<2\Delta$ and (b) $U>2\Delta$.}
    \label{schematic}
\end{figure}

Now that we are equipped with Eq.~\eqref{rhoF} to analytically obtain DOS
within a given order of strong coupling perturbation theory, and we have checked
that it gives reasonable results for the case of Mott transition in the Hubbard
model, let us break the sublattice symmetry by adding a scalar
potential $\pm\Delta$ to the two sublattices. This potential is known as ionic
potential and hence the Hamiltonian of the ionic Hubbard model is given by,
\bea
H&=&-t\sum_{i\in A,j\in B,\atop\sigma}(\cdag_{i\sigma}c_{j\sigma}+{\rm{h.c.}})+U\sum_i\ n_{i\uparrow}n_{i\downarrow}\nn\\
&+&\Delta\sum_{i\in A,\sigma}n_{i\sigma}-\Delta\sum_{j\in B,\sigma}n_{i\sigma}-\mu\sum_{i\sigma}n_{i\sigma}\label{ionic}
\eea
In atomic limit ($t=0$), the model reduces to classical Ising-like effective model 
that contains various insulating phases~\cite{HafezPRB}. At the simplest level, setting $t=0$ in 
the above Hamiltonian and corresponding to half-filling, the essential competition is between
$\Delta$ and $U$ terms. When the ionic potential dominates, i.e. $\Delta\gg U$ as can be seen
in the left part of the schematic drawing of Fig.~\ref{schematic}, 
both up and down spin electrons are pilled in a sublattice whose ionic potential is lower. 
In this  limit the $U$ is not enough to exclude the double occupancy.
However in the opposite limit of $U\gg \Delta$, the double occupancy is excluded and
the system becomes a Mott insulator with a charge gap $\sim U$. 
Then the important question is what is the nature of the ground state for $U\sim\Delta$ regime when the
fluctuations arising from the kinetic term ($t$) are turned on?

Tackling the problem with strong coupling perturbation theory, in this case the $H_0$ part of the
Hamiltonian will contain only $U$ and $\Delta$ terms and the perturbation term $H_1$ will be the
hopping term. Therefore we expect the method to be reliable only when both $\Delta$ and $U$ are 
quite larger than the hopping $t$. But even in the limit $U,\Delta\gg t$, it is 
interesting to have an idea of the nature of the gap when $U$ and $\Delta$ are comparable.

Now the $H_0$ part not only contains the parameter $U$, but also contains the energy scale $\Delta$.
Therefore the corresponding vertices of the auxiliary fermions have a build-in structure 
containing the competition between $U$ and $\Delta$. The role of ionic $\Delta$ can be easily 
incorporated as two different types of chemical potential for the two sublattices. If we denote
the self-energy of the auxiliary fermions on sublattice A and B with $\Gamma^{(A)}$ and $\Gamma^{(B)}$ respectively, we obtain the self-energy of auxiliary fermions on the lattice as:
\bea
 \Gamma(i\omega)=
 \begin{pmatrix}
 \Gamma^{(A)}(i\omega) & 0 \\ \\0 & \Gamma^{(B)}(i\omega)
 \end{pmatrix},
\eea
and DOS is given by
\bea
\rho(\omega)&=&-\frac{1}{\pi}\frac{1}{2}~\bigg(\sqrt{\frac{\Gamma^{(A)}(\omega)}{\Gamma^{(B)}(\omega)}}+\sqrt{\frac{\Gamma^{(B)}(\omega)}{\Gamma^{(A)}(\omega)}}\bigg)\nn\\
&\times&\lim_{\eta\rightarrow0^+}\sum_{\mathbf{k}} \mathrm{Im}\bigg[\bigg(\frac{1}{\frac{1}{\sqrt{\Gamma^{(A)}(\omega+i\eta)\Gamma^{(B)}(\omega+i\eta)}}-t|s(\mathbf{k})|}\nn\\
&&\quad\quad+\frac{1}{\frac{1}{\sqrt{\Gamma^{(A)}(\omega+i\eta)\Gamma^{(B)}(\omega+i\eta)}}+t|s(\mathbf{k})|}\bigg)\bigg]\label{dos-IHM}.
\eea 
Comparison between Eqns.~(\ref{dos-IHM}) and (\ref{graphene-dos}) leads to the following expression for DOS,
\bea
\rho(\omega)&=&\frac{1}{2}~\bigg(\sqrt{\frac{\Gamma^{(A)}(\omega)}{\Gamma^{(B)}(\omega)}}+\sqrt{\frac{\Gamma^{(B)}(\omega)}{\Gamma^{(A)}(\omega)}}\bigg)
\rho_0 ({\cal F}_1^{-1}),\nn\\
\label{graphenelikedos}
\eea
with
\be
{\cal F}_1={\sqrt{\Gamma^{(A)}(\omega)\Gamma^{(B)}(\omega)}}.
\label{rhoF1}
\ee

The representation Eq.~\eqref{graphenelikedos} is valid as long as
the function ${\cal F}_1$ is purely real. In the case of ionic Hubbard
model, the above function when evaluated at $\omega+i0^+$ is either purely
real that makes the above representation reliable, or purely imaginary.
In the later case, a more general formula for the green's function of 
hopping Hamiltonians derived by Horiguchi~\cite{Horiguchi} must be applied.
This has been summarized in Appendix~\ref{appendixdos}. The expression of
Horiguchi is valid for any complex argument. Evaluation of the resulting
DOS for purely imaginary arguments shows that it becomes identically zero.
Therefore the insulating gap is determined by 
\be
|{\cal F}_1^{-1}(0)| > 3 ~~~~\mbox{or}~~~\mbox{Re}\left({\cal F}_1(0)\right)=0
\label{gapcond2.eqn}
\ee

Note again that, so far we have not specified the
self-energy matrix elements $\Gamma^{(A)}$ and $\Gamma^{(B)}$ and therefore the discussion up to now 
remains quite general. Depending on the order of perturbation theory, these quantities may have
different expressions. But the important point is that their dynamical structure, as well as 
their parametric dependence on $U$ and $\Delta$ contains the essential physics of the 
interplay between the Mott insulating phase and band insulating phase. As before, 
the energy dependent quantity ${\cal F}_1$ determines the gap opening as well as the 
formation of Dirac nodes in the system.

Let us proceed with our discussion of the ionic Hubbard model by defining 
the mean occupation for a given spin projection on each sublattice at half-filling,
\bea
&&n^{(A)}=\frac{e^{\beta(u-\Delta)}+e^{-2\beta\Delta}}{1+2e^{\beta(u-\Delta)}+e^{-2\beta\Delta}},\nn\\
&&n^{(B)}=\frac{e^{\beta(u+\Delta)}+e^{2\beta\Delta}}{1+2e^{\beta(u+\Delta)}+e^{2\beta\Delta}},\label{occupation}
\eea
where for brevity we have used $u$ for $U/2$.
Note that in the case of simple Hubbard model where $\Delta=0$, the zero temperature
limit ($\beta\to\infty$) gives a very simple result $n^{(A)}=n^{(B)}=1/2$ for each spin 
projection.

\subsection{Zeroth order}
Now let us start by the lowest order of the perturbation theory for the ionic Hubbard model.
Keeping only zeroth order diagram in powers of $t$ depicted in Fig.~\ref{diagram},
the self-energies of the auxiliary fermions on two sublattices become:
\bea
\Gamma^{(A)}(i\omega)=\frac{1-n^{(A)}}{i\omega+U/2-\Delta}+\frac{n^{(A)}}{i\omega-U/2-\Delta}
\label{gammaA}\\
\Gamma^{(B)}(i\omega)=\frac{1-n^{(B)}}{i\omega+U/2+\Delta}+\frac{n^{(B)}}{i\omega-U/2+\Delta}
\label{gammaB}
\eea

\begin{figure}[tb]
    \includegraphics[width=8cm]{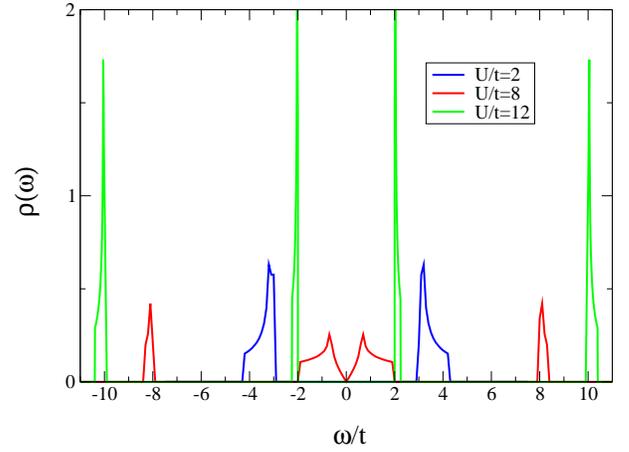}
    \caption{(Color online) DOS of the zeroth order diagram for $\Delta=4t$ in half-filled ionic Hubbard model on honeycomb lattice at zero temperature.
    The different colors as indicated in the legend correspond to band insulating state
    ($U=\Delta/2$), semimetallic state ($U=2\Delta$) and Mott insulating state ($U=3\Delta$).
    }
    \label{dos-ionic}
\end{figure}  

Let us first employ Eq.~\eqref{graphenelikedos} to generate a plot of DOS. As pointed out,
the present approach being a strong coupling expansion is reliable when $U,\Delta\gg t$. 
In generating the plots we set $\Delta=4t$ and $T=0$. As can be seen in Fig.~\ref{dos-ionic} 
we have three different situations. The blue plot corresponds to $U=2t=\Delta/2$ where there is a 
gap in the spectrum, and there are no signatures of upper and lower Hubbard bands. In this case
the gap is dominated by a single-particle character coming from the ionic potential $\Delta$.
By increasing $U$, we get to the red plot that corresponds to $U=8t=2\Delta$. There is a very 
beautiful linear V shaped pseudo-gap in the spectrum characteristic of a Dirac cone in two dimensions.
At the same time, there are also signatures of upper and lower Hubbard band formation at higher energy scales.
Upon further increase of the Hubbard parameter for $U=12t=3\Delta$ (green plot), again a gap 
opens up on top of a Dirac liquid state~\cite{Jafari2009}. This gap has a Mott nature and features 
of upper and lower Hubbard bands are visible. In table~\ref{table} we have extracted the precise
gap values from the criteria on ${\cal F}_1^{-1}$, Eq.~\eqref{gapcond2.eqn}.
\begin{table}[h]
\caption{The single-particle gap for the first two diagrams of Fig.~\ref{diagram} at $\Delta=4t$ and zero temperature. }
\begin{tabular}[t]{p{1.4cm} p{1.4cm} p{1.4cm} p{1.4cm}}
\hline%
$U/t$ & 2 & 8 & 12\\[0.7ex]
\hline
$E_{\rm gap}/t$ & 5.9558 & 0.0000 & 3.9998 \\[0.5ex]
\hline
\end{tabular}
\label{table}
\end{table}

Therefore the essential physics emerging here is that the competition between two gapped states
at $U\gg \Delta$ (Mott state) and $U\ll \Delta$ (Band insulating state) gives rise to a 
conducting state which in this case is a Dirac liquid state. This is in agreement with 
our previous DMFT finding~\cite{Ebrahimkhas}. 
However note that within the DMFT we find a conducting (Dirac) {\em region} sandwiched
between the Mott and band insulating phases, while in the present strong coupling
expansion the ensuing conducting (Dirac) state at the lowest order is a quantum critical Dirac state. 
Indeed the existence of 
a Dirac cone at $U=2\Delta$ can be seen analytically from the 
lowest order expressions \eqref{gammaA} and \eqref{gammaB}.
Let us first take the limit $T\to 0$ or equivalently $\beta\to\infty$. In this limit one has
$n^{(A)}=1/3$ and $n^{(B)}=2/3$ when both $U,\Delta >0$. Filpping the sign of $\Delta$ amounts
to swaping the occupancies of the two sublattices. In this limit the self-energies of the two
sublattices will become,
\be
   \Gamma^{(A/B)}(\omega)=\frac{1}{3}\left(\frac{2}{\omega}+\frac{1}{\omega\mp U} \right)
   ~~\mbox{for } U=2\Delta, T=0
\ee
The divergence of the above sublattice self-energies at $\omega=0$ makes ${\cal F}_1$ 
divergent at this point and therefore gives rise to vanishing DOS and hence a Dirac point.
Note that the existence of an intermediate Dirac phase which has been brought up with
state of the art DMFT, now can be seen analytically using even
a lowest order expression for the auxiliary fermion self-energies. Therefore the 
conducting phase that results from the competition between $U$ and $\Delta$ does
not seem to be artifact of infinite dimensions inherent in DMFT formulation.
Let us now go beyond the zeroth order and see how does the spectral gap evolves
upon going to higher orders of expansion.

\subsection{Beyond zeroth order}
Up to now, we have only considered the lowest order diagram of the Fig.~\ref{diagram}. 
Let us now add the second order diagram of Fig.~\ref{diagram}. 
The self-energy of auxiliary fermions of second order diagram on sublattice A  ($\Gamma^{(A)}_2$) for 
arbitrary temperature
is given in Appendix~\ref{appendixA}. The one for sublattice B is obtained by simply changing the
sign of $\Delta$, i.e. $\Delta \to -\Delta$. We have used subscript 2 in $\Gamma^{(A)}_2$ to stress 
that this self-energy is only related to second order diagram of Fig.~\ref{diagram}. Note that 
self-energy of auxiliary fermions on sublattice A in expansion up to second order is obtained by 
adding Eq.~\eqref{gamma2} to Eq.~\eqref{gammaA}. Also, the zero temperature limit of auxiliary fermion self-energies on two sublattices for second order diagram is presented in Appendix~\ref{appendixA}.
Having $\Gamma^{(A)}$ and $\Gamma^{(B)}$, we are able to calculate the single-particle gap. 
\begin{figure}[tb]
\includegraphics[width = 8.0cm]{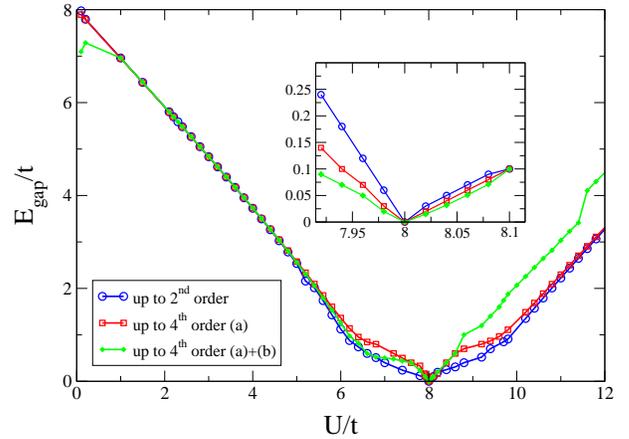}
\caption{(Color online) Calculated single-particle gap of the half-filled ionic Hubbard model on honeycomb lattice in zero temperature limit for $\Delta=4t$ up to second (blue circles), to fourth order diagram \ref{fourthorder}(a) (red squares) and to both fourth order diagrams of Fig.~\ref{fourthorder} (green diamonds). The inset zooms in the region around $U=2\Delta$.} 
\label{gap-IHM}
\end{figure}
The competition between interaction $U$ and ionic potential $\Delta$ at $\Delta=4t$ for zero 
temperature is shown in Fig.~\ref{gap-IHM}. This figure shows the value of gap as a function 
of $U$ for a fixed $\Delta=4t$ at zero temperature. The quantum critical metallic state
is at $U=8t$ that corresponds to $U=2\Delta$ where the gap entirely vanishes, and 
the spectrum of excitation contains a Dirac cone.
As pointed out earlier, the present strong coupling scheme being an expansion in powers of $t$
works better when the parameters satisfy $U,\Delta \gg t$. That is why we have chosen $\Delta=4t$
to address the competition between $U$ and $\Delta$ in presence of the hopping term.

As can be seen in Fig.~\ref{gap-IHM} (blue circles), in the presence of $\Delta$ for two diagrams of Fig.~\ref{diagram}, when the Hubbard interaction
term $U$ is small,  there is a gap in the spectrum of single-particle excitation. Since this gap
is continuously connected to $U\to 0$ limit, this gapped phase is a band insulating state.
When U increases, there is a critical point where the gap is zero, and the DOS is 
characterized by a Dirac cone around the $\omega=0$. As $U$ increases more, system enters Mott insulating phase.
It is important to see that for small values of $U/t$, although the parameters fall outside of the 
expected region of convergence of the present strong coupling approximation, the corresponding phases 
captured here is in agreement with our earlier studies using DMFT~\cite{Ebrahimkhas}.

\begin{figure}[tb]
\includegraphics[width = 4.5cm]{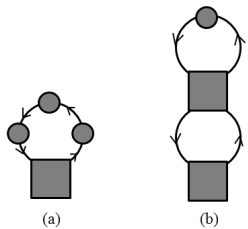}
\caption{Fourth order diagrams contributed to self-energy of auxiliary fermions.}
\label{fourthorder}
\end{figure}

In order to better treat the quantum fluctuations on top of the classical Hamiltonian $H_0$ of the 
ionic Hubbard model (i.e. $\Delta$ and $U$ terms involving commuting $n_{i\sigma}$ variables only), we consider higher orders in the perturbation theory. All fourth order diagrams are demonstrated in Fig. 18 of Ref.~\onlinecite{Senechal2000} but to illustrate their effect on the gap magnitude near the critical Dirac state $U=2\Delta$, 
we only consider two fourth order diagrams that depicted in Fig.~\ref{fourthorder}. Since for other fourth order diagrams, one needs to calculate three-particle connected correlation 
function which involves different expressions for $5!$ possible time orderings (one of the times 
can be set to zero) which makes it a formidable task to consider all of them. The self-energies of auxiliary fermions on sublattices $A/B$ in zero temperature limit for fourth order diagrams of Fig.~\ref{fourthorder}(a) $(\Gamma^{(A/B)}_{4(a)})$ and Fig.~\ref{fourthorder}(b) $(\Gamma^{(A/B)}_{4(b)})$ are given in Appendix~\ref{appendixB}.
The single-particle gaps obtained from adding diagram \ref{fourthorder}(a) (red squares) and both diagrams of Fig.~\ref{fourthorder} (green diamonds) to diagrams of Fig.~\ref{diagram} are shown in Fig.~\ref{gap-IHM}. As we see, by increasing the order of perturbation theory, the gap magnitudes for values of $U$ around $U=2\Delta$ 
become smaller (see inset of Fig.~\ref{gap-IHM}). However the present partial fourth order calculation is not
enough to imply that the quantum critical point at $U=2\Delta$ is broadened into a conducting (Dirac) 
region.

\section{discussions and summary\label{discussion}}
We have implemented a strong coupling expansion based on formalism proposed in Ref.~\onlinecite{Senechal98} 
on honeycomb lattice. We have used this method to study the semimetal to Mott insulator transition
on honeycomb lattice systems such as graphene, silicene. We have also used the ionic
Hubbard model to study the competition between
the ionic potential (mass term) and the Hubbard $U$. 

To study the influence of the on-site Coulomb interaction on honeycomb lattice, we have carried out the perturbative expansion of the auxiliary fermions around the atomic limit up to order $(t/U)^2$ and have analytically 
calculated the single-particle gap of the half-filled Hubbard model as function of $U$. The behavior
of a closed form function ${\cal F}(\omega)$ particularly at $\omega=0$ contains a great deal of information
about the possible interaction-induced gaps as well as about the Dirac nature of charge carriers 
on the honeycomb lattice. With this approach we find that the Mott transition for the Hubbard model
on the honeycomb lattice occurs at $2.38t$ which is in close agreement with previous studies. In Ref.~\onlinecite{SMIT1}, critical interaction $U_c\simeq 3t$ is found by slave-particle technique. Finite temperature cluster dynamical mean field theory with continuous time quantum Monte Carlo impurity solver anticipated $U_c \simeq 3.3t$ at zero temperature limit \cite{SMIT2}. Also, Seki and Ohta in Ref.~\onlinecite{SMIT3} within the variational cluster approximation showed that critical interaction for Mott transition is $\simeq 3.7t$.
Our result may be improved by going to higher orders of the perturbation theory. 

In the second part of the present paper we have studied the half-filled ionic Hubbard model on honeycomb 
lattice by strong coupling perturbation theory up to fourth order in terms of the hopping amplitude $t$. We have found the limits $U<2\Delta$ and $U>2\Delta$ 
are gapped states corresponding to band and Mott insulating phases, respectively. 
In the interaction strength $U=2\Delta$, owing to interplay between ionic potential and interaction, 
a semimetallic phase is restored. This agrees with earlier studies~\citep{HafezPRB,Ebrahimkhas}. 
It is interesting that the present result has been extracted within lowest, second and fourth order diagrams in terms of the behavior of function ${\cal F}_1$, particularly
around $\omega=0$. The detailed functional form of this function depends on the particular
order of the auxiliary fermion perturbation theory. 

The present study can be directly relevant to recent graphene/SiC where a gap of $0.5$ eV has
been found~\cite{Conrad}. In this case the gap of $0.5$ eV is jointly determined by a single-particle
gap parameter $\Delta$ and the many-particle (Mott) gap parameter $U$. 

The present strong coupling scheme seems to give reasonable results about the nature of the
gap in the spectrum of excitation. The method seems to be capable of unbiased estimate of the 
excitation spectrum in strongly correlated systems.

\section{acknowledgements}
E.A. was supported by the National Elite Foundation of Iran.
S.A.J. was supported by the Alexander von Humboldt foundation, Germany.

\appendix
\section{Exact expression for DOS on honeycomb lattice \label{appendixdos}}
We are going to represent the DOS of arbitrary complex frequency $\xi$ on honeycomb lattice. 
According to Ref. \onlinecite{Horiguchi}, the DOS for tight-binding model on the honeycomb
lattice can be expressed as,
\bea
\rho &=&-\frac{1}{\pi}~\sum_{\mathbf{k}} \mathrm{Im}\ \mathrm{Tr}\ G(\xi,\mathbf{k}),\nn\\
&=&-\frac{1}{\pi}\ \mathrm{Im}\bigg(2\ \xi\ G_\xi(\frac{1}{2}(\xi^2-3);0,0)\bigg)\label{doshoriguchi},
\eea   
where 
\bea
G_\xi(\xi;0,0)=\frac{1}{2\pi}\ g\ \tilde{K}(\mathit{k}),
\eea
is the local green's function evaluated at general complex argument $\xi$
 \begin{figure}[tb]
\includegraphics[width =8cm]{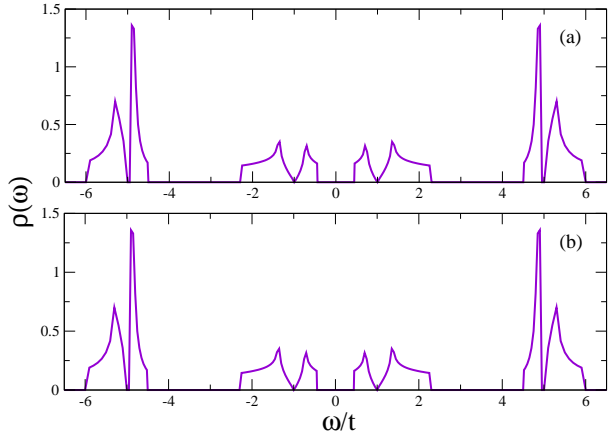}
\caption{DOS of the half-filled ionic Hubbard model on honeycomb lattice for $U=6$, $\Delta=2$ and $T=0$ up to order $t^2$ obtained from (a) Eq. (\ref{doshoriguchi}) for $\eta=10^{-4}$ and (b) Eq.~(\ref{doslikegraphene}).}
\label{figappendix}
\end{figure}
and $g$, $\mathit{k}$ and $\tilde{K}(\mathit{k})$ are given as follows:
\bea
&&g=\frac{2}{((2\xi+3)^{1/2}-1)^{3/2}((2\xi+3)^{1/2}-1)^{1/2}},\\
&&\mathit{k}=\frac{4(2\xi+3)^{1/4}}{((2\xi+3)^{1/2}-1)^{3/2}((2\xi+3)^{1/2}-1)^{1/2}},
\eea

\bea
\tilde{K}(\mathit{k})=\left\lbrace
\begin{array}{cccc}
K(\mathit{k})  & \mathrm{Im}\xi>0\ \mathrm{and} \ \mathrm{Im}\mathit{k}<0\\
&\ \mathrm{or}\ \mathrm{Im}\xi<0\  \mathrm{and}\  \mathrm{Im}\mathit{k}>0,\\
\\
K(\mathit{k})+2iK^\prime({\mathit{k}}) & \mathrm{Im}\xi>0\  \mathrm{and}\  \mathrm{Im}\mathit{k}>0,\\
\\
K(\mathit{k})-2iK^\prime({\mathit{k}}) & \mathrm{Im}\xi<0\  \mathrm{and}\  \mathrm{Im}\mathit{k}<0,
\end{array} \right.
\eea
where $K(\mathit{k})$ and $K^\prime({\mathit{k}})$ are the complete elliptic integral of the first kind and the complete elliptic integral of the first kind with complementary modulus of $\mathit{k}$, respectively.

The above formula is quite general. However in the $\eta\to 0$ limit it turns out that 
when $\xi=\mathtt{F}(\omega+i\eta)$, DOS (\ref{doshoriguchi}) reduces to,

\bea
\rho(\omega)=\frac{|\mathtt{F}(\omega)|}{\pi^2}\frac{1}{\sqrt{Z_0}}\ K\bigg(\sqrt{\frac{Z_1}{Z_0}}\bigg)\label{doslikegraphene},
\eea
where $Z_0$ and $Z_1$ are those introduced in Eqns. (\ref{z0}) and (\ref{z1}), albeit substitute $\omega$ with $\mathtt{F}(\omega)$. If we assume $\xi$ in Eq.~\eqref{doshoriguchi} has infinitismal imaginary part, 
$\omega+i0^+$, the resulting DOS will be DOS of graphene.
Also, DOS for ionic Hubbard model on honeycomb lattice for $U=6$, $\Delta=2$ and $T=0$ obtained from Eq.~(\ref{doshoriguchi}) and Eq.~(\ref{doslikegraphene}) are shown on Figs. \ref{figappendix}(a) and \ref{figappendix}(b), respectively. As we can see, the two DOS well coincide, demonstrating that the above representation
works well for situations where $\mathtt {F}$ is purely imaginary or purely real.

\begin{widetext}
\section{Dual fermion self-energies in ionic Hubbard model}
This Appendix gives the calculated self-energies of auxiliary fermions in second and fourth order.
\subsection{Second order\label{appendixA}}
 
Introducing the  following definitions,
\bea
Z^{(A)}=1+2 e^{\beta(u-\Delta)}+e^{-2\beta\Delta},
~~~~~n_F(x)=\frac{1}{e^{\beta x}+1},
~~~~~n_B(x)=\frac{1}{e^{\beta x}-1},
\eea
the self-energy of the second order diagram of Fig.~\ref{diagram} on sublattice $A$ ($\Gamma^{(A)}_2$) at 
arbitrary temperature $1/\beta$ reads:
\bea
\Gamma^{(A)}_2(i\omega)&=&\frac{-1.15\ t^2\ (2n^{(A)}-1)\ n_F(\Delta+u)}{(i\omega-\Delta)^2-u^2}\ \bigg(\frac{1-n^{(B)}}{\Delta+u}+\frac{n^{(B)}}{\Delta} \bigg)\nn\\
&+&\frac{2.3\ t^2\ (2n^{(A)}-1)\ n_B(2\Delta)}{(i\omega-\Delta)^2-u^2}\ \bigg(\frac{1-n^{(B)}}{i\omega-3\Delta-u}+\frac{n^{(B)}}{i\omega-3\Delta+u} \bigg)\nn\\
&-&\frac{1.15\ t^2}{(i\omega-\Delta)^2-u^2}\ \bigg(\beta u^2 n^{(A)} (1-n^{(A)})+\frac{\beta u^2}{({Z^{(A)}})^2} (e^{-2\beta\Delta}-e^{2\beta(u-\Delta)})+u(1-n^{(A)})\bigg)\nn\\
&&\times \bigg\lbrace(1-n^{(B)})\bigg(\frac{n_F(\Delta+u)}{u(\Delta+u)}+\frac{n_F(-\Delta-u)}{\Delta(\Delta+u)}-\frac{n_F(\Delta-u)}{\Delta u} \bigg)+n^{(B)}\bigg(\frac{n_F(\Delta+u)}{\Delta u}-\frac{n_F(\Delta-u)}{u(\Delta-u)}-\frac{n_F(-\Delta+u)}{\Delta(-\Delta+u)}  \bigg)\bigg\rbrace\nn\\
&+&\frac{2.3\ t^2\ u\ (1-n^{(A)})}{(i\omega-\Delta)^2-u^2}\ \bigg\lbrace (1-n^{(B)})\bigg( \frac{n_F(-\Delta-u)-n_F(\Delta-u)}{4\Delta^2}+\frac{-\beta\ n_F(\Delta-u)+\beta\ (n_F(\Delta-u))^2}{2\Delta}\bigg)\nn\\
&&\qquad\qquad\qquad\qquad\qquad\quad+n^{(B)}\bigg( \frac{n_F(-\Delta+u)-n_F(\Delta-u)}{4(\Delta-u)^2}+\frac{-\beta\ n_F(\Delta-u)+\beta\ (n_F(\Delta-u))^2}{2(\Delta-u)}\bigg)\bigg\rbrace\nn\\
&+&\frac{1.15\ t^2\ \Big((i\omega-\Delta)(2n^{(A)}-1)+u\Big)}{(i\omega-\Delta)^2-u^2}\bigg\lbrace (1-n^{(B)})\bigg( \frac{n_F(-\Delta-u)-n_F(\Delta+u)}{4(\Delta+u)^2}+\frac{-\beta\ n_F(\Delta+u)+\beta\ (n_F(\Delta+u))^2}{2(\Delta+u)}\bigg)\nn\\
&&\qquad\qquad\qquad\qquad\qquad\quad+n^{(B)}\bigg( \frac{n_F(-\Delta+u)-n_F(\Delta+u)}{4\Delta^2}+\frac{-\beta\ n_F(\Delta+u)+\beta\ (n_F(\Delta+u))^2}{2\Delta}\bigg)\bigg\rbrace\nn\\
&+&\frac{4.6\ t^2\ u^2}{((i\omega-\Delta)^2-u^2)^2}\ \bigg(\frac{1-n^{(B)}}{i\omega+\Delta+u}+\frac{n^{(B)}}{i\omega+\Delta-u} \bigg)\ \bigg(n^{(A)}(1-n^{(A)})+\frac{e^{\beta(u-\Delta)}}{Z^{(A)}} \bigg)\nn\\
&+&1.15\ t^2\ u\ \bigg(\frac{(1-n^{(A)})}{2\ (i\omega-\Delta+u)^2}+\frac{1}{4\ (i\omega-\Delta-u)^2}\bigg)\nn\\ &\times &\bigg\lbrace(1-n^{(B)})\bigg(\frac{n_F(\Delta+u)}{u(\Delta+u)}+\frac{n_F(-\Delta-u)}{\Delta(\Delta+u)}-\frac{n_F(\Delta-u)}{\Delta u} \bigg)+n^{(B)}\bigg(\frac{n_F(\Delta+u)}{\Delta u}-\frac{n_F(\Delta-u)}{u(\Delta-u)}-\frac{n_F(-\Delta+u)}{\Delta(-\Delta+u)}  \bigg)\bigg\rbrace\nn\\
&+&\frac{1.15\ t^2\ (2n^{(A)}-1)}{4\ (i\omega-\Delta-u)^2}\ \bigg\lbrace(1-n^{(B)})\bigg(\frac{n_F(\Delta-u)}{\Delta}+\frac{n_F(\Delta+u)}{\Delta+u}-\frac{(2\Delta+u)\ n_F(-\Delta-u)}{\Delta (\Delta+u)} \bigg)\nn\\
&&\qquad\qquad\qquad\qquad\qquad\qquad\qquad +n^{(B)}\bigg(\frac{n_F(\Delta-u)}{\Delta-u}+\frac{n_F(\Delta+u)}{\Delta}+\frac{(2\Delta-u)\ n_F(-\Delta+u)}{\Delta(-\Delta+u)}  \bigg)\bigg\rbrace\nn\\
&-&\frac{1.15\ t^2\ (2 n^{(A)}-1)}{(i\omega-\Delta-u)^2}\ \bigg\lbrace \frac{(1-n^{(B)})\Big(n_F(-\Delta-u)+n_B(2\Delta) \Big)}{i\omega-3\Delta-u}+\frac{n^{(B)}\Big(n_F(-\Delta+u)+n_B(2\Delta) \Big)}{i\omega-3\Delta+u}\bigg\rbrace\nn\\
&-&\frac{1.15\ t^2\ (2n^{(A)}-1)}{(i\omega-\Delta+u)^2}\bigg\lbrace  n_B(2\Delta)\bigg(\frac{1-n^{(B)}}{i\omega-3\Delta-u}+\frac{n^{(B)}}{i\omega-3\Delta+u} \bigg)\nn\\
&&\qquad\qquad\qquad\qquad\qquad\qquad-n_F(\Delta+u)\bigg(\frac{n^{(B)}(i\omega+\Delta+u)}{4\Delta^2}+\frac{(1-n^{(B)})(i\omega+\Delta+3u)}{4(\Delta+u)^2} \bigg)\bigg\rbrace\nn\\
&+&\frac{1.15\ t^2\ (2n^{(A)}-1)}{2\ (i\omega-\Delta+u)}\bigg(\beta\ n_F(\Delta+u)-\beta\ (n_F(\Delta+u))^2 \bigg)\ \bigg(\frac{1-n^{(B)}}{\Delta+u}+\frac{n^{(B)}}{\Delta} \bigg)\nn\\
&-&\frac{1.15\ t^2\ (2n^{(A)}-1)\ n^{(B)}\ n_F(-\Delta+u)}{i\omega-3\Delta+u}\ \bigg(\frac{1}{4\Delta^2}-\frac{1}{\Delta\ (i\omega-\Delta-u)} \bigg)\nn\\
&-&\frac{1.15\ t^2\ (2n^{(A)}-1)\ (1-n^{(B)})\ n_F(-\Delta-u)}{i\omega-3\Delta-u}\ \bigg(\frac{1}{4(\Delta+u)^2}-\frac{1}{(\Delta+u)\ (i\omega-\Delta-u)} \bigg)\label{gamma2},
\eea
where $u=U/2$ and $n^{(A)}$, $n^{(B)}$ are given in Eq.~(\ref{occupation}).
By flipping the sign of $\Delta$ in the self-energy $\Gamma^{(A)}_2$ of sublattice A, 
one can obtain the self-energy of auxiliary fermions on sublattice B in given order ($\Gamma^{(B)}_2$).
Taking the zero temperature limit, the second order self-energy of auxiliary fermions on sublattice A and B 
is simplified to,
\bea
\Gamma_2^{(A)}(i\tilde{\omega})=\left\lbrace
\begin{array}{cccc}
\frac{1.15 \tilde{t}^2}{\Delta}\Bigg[\frac{-\tilde{u}^2}{2(\tilde{u}^2-1)^2((i\tilde{\omega}-1)^2-\tilde{u}^2)}+\frac{3     \tilde{u}^2 (i\tilde{\omega}+1)}{((i\tilde{\omega}-1)^2-\tilde{u}^2)^2 ((i\tilde{\omega}+1)^2-\tilde{u}^2)}
+\frac{1}{4(\tilde{u}^2-1)}\big( \frac{1}{(i\tilde{\omega}+\tilde{u}-1)}+\frac{1}{(i\tilde{\omega}-\tilde{u}-1)}\big)^2\Bigg] &  \tilde{u} >1 \\
\\
\infty & \tilde{u}=1\\
\\
\frac{1.15 \tilde{t}^2}{\Delta}\Bigg[ \frac{\tilde{u}}{((i\tilde{\omega}-1)^2-\tilde{u}^2)}\big(\frac{1}{2(1-\tilde{u})^2}-\frac{1}{(1-\tilde{u})}\big)-\frac{1}{4 (i\tilde{\omega}+\tilde{u}-1)}+\frac{1}{2(1-\tilde{u})}\big(\frac{1}{(i\tilde{\omega}-\tilde{u}-1)^2}+\frac{\tilde{u}}{(i\tilde{\omega}+\tilde{u}-1)^2}\big)\\
 \\
 +\frac{1}{(i\tilde{\omega}+\tilde{u}-3)}\big(\frac{1}{(i\tilde{\omega}-\tilde{u}-1)}-\frac{1}{2}\big)^2\Bigg] & \tilde{u} < 1
\end{array} \right.
\eea
and
\bea
\Gamma_2^{(B)}(i\tilde{\omega})=\left\lbrace
\begin{array}{cccc}
\frac{1.15 \tilde{t}^2}{\Delta}\Bigg[\frac{\tilde{u}^2}{2(\tilde{u}^2-1)^2((i\tilde{\omega}+1)^2-\tilde{u}^2)}+\frac{3 \tilde{u}^2 (i\tilde{\omega}-1)}{((i\tilde{\omega}+1)^2-\tilde{u}^2)^2 ((i\tilde{\omega}-1)^2-\tilde{u}^2)}
-\frac{1}{4(\tilde{u}^2-1)}\big( \frac{1}{(i\tilde{\omega}+\tilde{u}+1)}+\frac{1}{(i\tilde{\omega}-\tilde{u}+1)}\big)^2\Bigg] &   \tilde{u} >1 \\
\\
\infty & \tilde{u}=1\\
\\
\frac{1.15 \tilde{t}^2}{\Delta}\Bigg[\frac{-1}{(\tilde{u}-1)((i\tilde{\omega}+1)^2-\tilde{u}^2)}-\frac{1}{4(\tilde{u}-1)^2(i\tilde{\omega}-\tilde{u}+1)}
+\frac{1}{2(\tilde{u}-1)(i\tilde{\omega}-\tilde{u}+1)^2}
+\frac{(i\tilde{\omega}+3\tilde{u}-1)}{4(\tilde{u}+1)^2(i\tilde{\omega}+\tilde{u}+1)^2}
\\
\\+\frac{1}{(i\tilde{\omega}-\tilde{u}+3)}\big(\frac{1}{(i\tilde{\omega}-\tilde{u}+1)}-\frac{1}{(i\tilde{\omega}+\tilde{u}+1)}\big)^2\Bigg] & \tilde{u} < 1
\end{array} \right.
\eea
where we have used dimensionless quantities $\tilde{t}=\frac{t}{\Delta}, \tilde{\omega}=\frac{\omega}{\Delta}$ 
and $\tilde{u}=\frac{u}{\Delta}$. It is interesting to note that the above expressions have an overall 
scale $\tilde t^2/\Delta$ multiplied by a function of $\omega/\Delta$ and $u/\Delta$. This scaling functional
form continues to higher order as we see in the next subsection.

\subsection{Fourth order\label{appendixB}}
We have undertaken the cumbersome task of calculation of two of the fourth order diagrams discussed in 
the text. The arbitrary temperature expression for the fourth order contributions is huge. Therefore
in this Appendix we only report their zero temperature limit. 
The zero temperature limit of self-energies of auxiliary fermions on sublattices $A/B$ corresponding
to diagram~\ref{fourthorder}(a) $(\Gamma^{(A/B)}_{4(a)})$ and~\ref{fourthorder}(b) $(\Gamma^{(A/B)}_{4(b)})$ 
are separately calculated in two regions $u > \Delta (\tilde u >1)$ and $u < \Delta (\tilde u <1)$. 
This separation naturally arises when we take the zero temperature limit. 

In $\tilde{u}>1$ limit, the self-energies are given by:
\bea
\Gamma^{(A)}_{4(a)}(i\tilde{\omega})&=&\frac{\tilde{t}^4}{\Delta}\Bigg[ \frac{3\tilde{u}^2}{8((i\tilde{\omega}-1)^2-\tilde{u}^2)^2}\bigg(\frac{1}{i\tilde{\omega}+\tilde{u}+1}+\frac{1}{i\tilde{\omega}-\tilde{u}+1}\bigg)^2\bigg(\frac{1}{i\tilde{\omega}+\tilde{u}-1}+\frac{1}{i\tilde{\omega}-\tilde{u}-1}\bigg)\nn\\
&+&\frac{\tilde{u}}{8}\bigg(\frac{1}{(i\tilde{\omega}+\tilde{u}-1)^2}+\frac{1}{(i\tilde{\omega}-\tilde{u}-1)^2}\bigg)\bigg(\frac{(\tilde{u}+1)^2+1}{8\tilde{u}(\tilde{u}+1)^3}-\frac{\tilde{u}+2}{8(\tilde{u}+1)^2}+\frac{1}{8\tilde{u}^2(1-\tilde{u})^3}+\frac{\tilde{u}-1}{8\tilde{u}^2}\bigg)\nn\\
&+&\frac{\tilde{u}}{8((i\tilde{\omega}-1)^2-\tilde{u}^2)}\bigg(-\frac{\tilde{u}+2}{8\tilde{u}(\tilde{u}+1)^2}+\frac{\tilde{u}-2}{8\tilde{u}(\tilde{u}-1)^2}+\frac{5\tilde{u}-2}{16\tilde{u}(\tilde{u}-1)^4}+\frac{2\tilde{u}^2+7\tilde{u}+3}{16(\tilde{u}+1)^3}+\frac{2-\tilde{u}}{16\tilde{u}(\tilde{u}+1)^4}+\frac{2\tilde{u}^2-3\tilde{u}+7}{16(1-\tilde{u})^3}\bigg)\Bigg],\nn\\
\eea

\bea
\Gamma^{(B)}_{4(a)}(i\tilde{\omega})&=&\frac{\tilde{t}^4}{\Delta}\Bigg[\frac{3\tilde{u}^2}{8((i\tilde{\omega}+1)^2-\tilde{u})^2}\bigg(\frac{1}{i\tilde{\omega}+\tilde{u}-1}+\frac{1}{i\tilde{\omega}-\tilde{u}-1}\bigg)^2\bigg(\frac{1}{i\tilde{\omega}+\tilde{u}+1}+\frac{1}{i\tilde{\omega}-\tilde{u}+1}\bigg)\nn\\
&+&\frac{\tilde{u}}{8}\bigg(\frac{1}{(i\tilde{\omega}+\tilde{u}+1)^2}+\frac{1}{(i\tilde{\omega}-\tilde{u}+1)^2}\bigg)\bigg(\frac{(\tilde{u}-1)^2+1}{8\tilde{u}(\tilde{u}-1)^3}+\frac{\tilde{u}-2}{8(\tilde{u}-1)^2}+\frac{1}{8\tilde{u}^2(\tilde{u}+1)^3}-\frac{\tilde{u}+1}{8\tilde{u}^2}\bigg)\nn\\
&+&\frac{\tilde{u}}{8((i\tilde{\omega}+1)^2-\tilde{u}^2)}\bigg(\frac{2-\tilde{u}}{8\tilde{u}(\tilde{u}-1)^2}+\frac{\tilde{u}+2}{8\tilde{u}(\tilde{u}+1)^2}+\frac{5\tilde{u}+2}{16\tilde{u}(\tilde{u}+1)^4}-\frac{2\tilde{u}^2-7\tilde{u}+3}{16(\tilde{u}-1)^3}-\frac{\tilde{u}+2}{16\tilde{u}(\tilde{u}-1)^4}+\frac{2\tilde{u}^2+3\tilde{u}+7}{16(\tilde{u}+1)^3}\bigg)\Bigg],\nn\\
\eea

\bea
\Gamma^{(A)}_{4(b)}(i\tilde{\omega})&=&\frac{(1.15)^2\tilde{t}^4}{\Delta}\Bigg[-\frac{9\ \tilde{u}^4}{4((i\tilde{\omega}-1)^2-\tilde{u}^2)^2\ ((i\tilde{\omega}+1)^2-\tilde{u}^2)^2} \bigg(\frac{1}{i\tilde{\omega}+\tilde{u}-1}+\frac{1}{i\tilde{\omega}-\tilde{u}-1}\bigg)\nn\\
&+&\frac{\tilde{u}^2}{8 (\tilde{u}^2-1)\ ((i\tilde{\omega}-1)^2-\tilde{u}^2)^2}\bigg( \frac{1}{i\tilde{\omega}+\tilde{u}+1}-\frac{1}{i\tilde{\omega}-\tilde{u}+1}\bigg)^2-\frac{\tilde{u}^4}{4\ (\tilde{u}^2-1)^2\ ((i\tilde{\omega}-1)^2-\tilde{u}^2)^2\  ((i\tilde{\omega}+1)^2-\tilde{u}^2)}\nn\\
&-&\frac{(4\tilde{u}^3-3\tilde{u}^2-2\tilde{u}+1)}{64\ (\tilde{u}^2-1)^2\ (\tilde{u}-1)^2}\bigg( \frac{1}{i\tilde{\omega}+\tilde{u}-1}-\frac{1}{i\tilde{\omega}-\tilde{u}-1}\bigg)^2\nn\\
&+&\frac{3}{128}\bigg( \frac{\tilde{u}^3+4\tilde{u}^2+6\tilde{u}+2}{(1+\tilde{u})^3}-\frac{\tilde{u}^2(\tilde{u}-2)}{(\tilde{u}-1)^3} \bigg)\bigg( \frac{1}{(i\tilde{\omega}+\tilde{u}-1)^2}+\frac{1}{(i\tilde{\omega}-\tilde{u}-1)^2}\bigg)\nn\\
&+&\frac{1}{2 ((i\tilde{\omega}-1)^2-\tilde{u}^2)}\bigg( \frac{3(3\tilde{u}+2)}{128}-\frac{3\tilde{u}^3(3\tilde{u}^2+4)}{128(\tilde{u}-1)^4}+\frac{(1+\tilde{u}^2)(\tilde{u}^4+2\tilde{u}^2-1)}{16(\tilde{u}^2-1)^4}+\frac{3\tilde{u}^2}{32 (\tilde{u}-1)^2}\nn\\
&&\qquad\qquad-\frac{3(\tilde{u}^4+5\tilde{u}^3+10\tilde{u}^2+3\tilde{u})}{64 (\tilde{u}+1)^4}\bigg)\Bigg],\nn\\
\eea

\bea
\Gamma^{(B)}_{4(b)}(i\tilde{\omega})&=&\frac{(1.15)^2\tilde{t}^4}{\Delta}\Bigg[\frac{-9\tilde{u}^4}{4((i\tilde{\omega}+1)^2-\tilde{u}^2)^2((i\tilde{\omega}-1)^2-\tilde{u}^2)^2}\bigg(\frac{1}{i\tilde{\omega}+\tilde{u}+1}+\frac{1}{i\tilde{\omega}-\tilde{u}+1}\bigg)\nn\\
&+&\frac{\tilde{u}^2}{8(1-\tilde{u}^2)((i\tilde{\omega}+1)^2-\tilde{u}^2)^2}\bigg(\frac{1}{i\tilde{\omega}+\tilde{u}-1}-\frac{1}{i\tilde{\omega}-\tilde{u}-1}\bigg)^2\nn\\
&+&\frac{\tilde{u}^4}{4(\tilde{u}^2-1)^2((i\tilde{\omega}^2+1)^2-\tilde{u}^2)^2 ((i\tilde{\omega}-1)^2-\tilde{u}^2)}-\frac{(4\tilde{u}^3+3\tilde{u}^2-2\tilde{u}-1)}{64(1+\tilde{u})^2(\tilde{u}^2-1)^2}\bigg( \frac{1}{i\tilde{\omega}+\tilde{\omega}+1}-\frac{1}{i\tilde{\omega}-\tilde{u}+1}\bigg)^2\nn\\
&+&\frac{3}{128}\bigg(\frac{\tilde{u}^2(\tilde{u}+2)}{(\tilde{u}+1)^3}-\frac{\tilde{u}^3-4\tilde{u}^2+6\tilde{u}-2}{(\tilde{u}-1)^3}\bigg)\bigg(\frac{1}{(i\tilde{\omega}-\tilde{u}+1)^2}+\frac{1}{(i\tilde{\omega}+\tilde{u}+1)^2}\bigg)\nn\\
&-&\frac{1}{2((i\omega+1)^2-\tilde{u}^2)}\bigg(\frac{-3(3\tilde{u}-2)}{128}+\frac{3\tilde{u}^3(3\tilde{u}^2+4)}{128(\tilde{u}+1)^4}+\frac{(\tilde{u}^2+1)(\tilde{u}^4+2\tilde{u}^2+1)}{16(\tilde{u}^2-1)^4}+\frac{3\tilde{u}^2}{32(\tilde{u}+1)^2}\nn\\
&&\qquad\qquad-\frac{3(\tilde{u}^4-5\tilde{u}^3+10\tilde{u}^2-3\tilde{u})}{64(\tilde{u}-1)^4}\bigg)\Bigg].\nn\\
\eea
In the $\tilde{u}<1$, the self-energies are expressed as,
\bea
\Gamma^{(A)}_{4(a)}(i\tilde{\omega})&=&\frac{-\tilde{t}^4}{\Delta}\Bigg[\frac{1}{(i\tilde{\omega}-\tilde{u}-1)^2}\bigg(\frac{1}{2(\tilde{u}-1)(i\tilde{\omega}+\tilde{u}-3)^2}+\frac{1}{4(\tilde{u}-1)^2(i\tilde{\omega}+\tilde{u}-3)}+\frac{1}{16(\tilde{u}-1)}+\frac{(\tilde{u}-3)(\tilde{u}+1)}{16(\tilde{u}-1)^3}\bigg)\nn\\
&-&\frac{2}{(i\tilde{\omega}-\tilde{u}-1)(i\tilde{\omega}+\tilde{u}-3)}\bigg(\frac{1}{4(\tilde{u}-1)(i\tilde{\omega}+\tilde{u}-3)}-\frac{\tilde{u}-2}{8(\tilde{u}-1)^2}\bigg)+\frac{\tilde{u}}{((i\tilde{\omega}-1)^2-\tilde{u}^2)}\bigg(\frac{3}{8(\tilde{u}-1)^4}+\frac{4(3-\tilde{u})}{16(\tilde{u}-1)^3}\bigg)\nn\\
&+&\frac{2\tilde{u}(\tilde{u}-3)}{16(\tilde{u}-1)^3(i\tilde{\omega}+\tilde{u}-1)^2}+\frac{2\tilde{u}-3}{16(\tilde{u}-1)^2(i\tilde{\omega}+\tilde{u}-1)}+\frac{3-2\tilde{u}}{16(\tilde{u}-1)^2(i\tilde{\omega}+\tilde{u}-3)}+\frac{1}{8(\tilde{u}-1)(i\tilde{\omega}+\tilde{u}-3)^2}\Bigg],\nn\\
\eea

\bea
\Gamma^{(B)}_{4(a)}&=&\frac{-\tilde{t}^4}{\Delta}\Bigg[\frac{1}{(i\tilde{\omega}+\tilde{u}+1)}\bigg(\frac{-1}{(i\tilde{\omega}-\tilde{u}+1)^2(i\tilde{\omega}-\tilde{u}+3)^2}+\frac{1}{4(\tilde{u}-1)^2(i\tilde{\omega}-\tilde{u}+1)^2}-\frac{1}{(i\tilde{\omega}+\tilde{u}+1)^2(i\tilde{\omega}-\tilde{u}+3)^2}\nn\\
&&\qquad\qquad+\frac{1}{4(\tilde{u}-1)^2(i\tilde{\omega}+\tilde{u}+1)^2}+\frac{3}{16(\tilde{u}-1)^4}\bigg)\nn\\
&+&\frac{1}{i\tilde{\omega}-\tilde{u}+1}\bigg(\frac{2}{(i\tilde{\omega}+\tilde{u}+1)^2(i\tilde{\omega}-\tilde{u}+3)^2}-\frac{1}{2(\tilde{u}-1)^2(i\tilde{\omega}+\tilde{u}+1)^2}-\frac{1}{2(\tilde{u}-1)^3(i\tilde{\omega}+\tilde{u}+1)}-\frac{3}{16(\tilde{u}-1)^4}\bigg)\nn\\
&+&\frac{1}{8(\tilde{u}-1)^2(i\tilde{\omega}-\tilde{u}+1)^2}\Bigg],
\eea

\bea
\Gamma^{(A)}_{4(b)}(i\tilde{\omega})&=&\frac{(1.15)^2\tilde{t}^4}{\Delta}\Bigg[ \frac{1}{(i\tilde{\omega}-\tilde{u}-1)^2}\bigg(\frac{1}{8(i\tilde{\omega}-\tilde{u}-5)}(1+\frac{1}{\tilde{u}-1})^2+\frac{(1+\tilde{u})^2-3(\tilde{u}^2-1)}{16\tilde{u}(\tilde{u}-1)(i\tilde{\omega}+\tilde{u}-3)}-\frac{3\tilde{u}-2}{4\tilde{u}(\tilde{u}-1)^2(i\tilde{\omega}-\tilde{u}-3)}\nn\\
&&\qquad\qquad\ \ -\frac{1}{4\tilde{u}(i\tilde{\omega}+\tilde{u}-1)}-\frac{1}{4(i\tilde{\omega}+\tilde{u}-3)^2}+\frac{1}{2(\tilde{u}-1)(i\tilde{\omega}-\tilde{u}-3)^2}\bigg)\nn\\
&+&\frac{1}{(i\tilde{\omega}-\tilde{u}-1)}\bigg( \frac{2\tilde{u}-1}{2\tilde{u}(\tilde{u}^2-1)^2(i\tilde{\omega}-\tilde{u}-3)}+\frac{\tilde{u}-2}{4\tilde{u}(\tilde{u}-1)(i\tilde{\omega}+\tilde{u}-3)}+\frac{2-\tilde{u}^2}{16(\tilde{u}-1)^2(i\tilde{\omega}+\tilde{u}-5)}\nn\\
&&\qquad\qquad-\frac{1}{2(\tilde{u}^2-1)(i\tilde{\omega}-\tilde{u}-3)^2}+\frac{1}{8(i\tilde{\omega}+\tilde{u}-3)^2}\bigg)\nn\\
&+&\frac{1}{128(i\tilde{\omega}+\tilde{u}-5)}(1+\frac{1}{\tilde{u}-1})^2+\frac{\tilde{u}+(\tilde{u}-2)^2}{32\tilde{u}(\tilde{u}-1)(i\tilde{\omega}+\tilde{u}-3)} -\frac{(\tilde{u}-1)^2+\tilde{u}+1}{16\tilde{u}(1+\tilde{u})(\tilde{u}^2-1)^2(i\tilde{\omega}-\tilde{u}-3)}\nn\\
&-&\frac{1}{16(i\tilde{\omega}+\tilde{u}-3)^2}\frac{1}{8(\tilde{u}-1)(1+\tilde{u})^2(i\tilde{\omega}-\tilde{u}-3)^2}\nn\\
&+&\frac{\tilde{u}}{((i\tilde{\omega}-1)^2-\tilde{u}^2)}\bigg( \frac{(\tilde{u}-1)^3+2\tilde{u}}{4\tilde{u}(\tilde{u}^2-1)^2}+\frac{\tilde{u}+(\tilde{u}-1)^2}{16(\tilde{u}-1)(\tilde{u}-2)^2}+\frac{1}{4\tilde{u}(1-\tilde{u})}+\frac{2\tilde{u}^2-3\tilde{u}+1}{16(\tilde{u}-1)^4}+\frac{1-4\tilde{u}}{16(\tilde{u}-1)^2}\bigg)\nn\\
&+&\frac{1}{(i\tilde{\omega}+\tilde{u}-1)}\bigg( \frac{-9}{128}-\frac{1}{128(\tilde{u}-1)^2}+\frac{1}{8(\tilde{u}-1)(1+\tilde{u})^3}+\frac{3\tilde{u}-4}{128\tilde{u}(\tilde{u}-1)}-\frac{(\tilde{u}+1)^2+(\tilde{u}-1)(\tilde{u}+2)}{32(\tilde{u}^2-1)^2}\bigg)\nn\\
&+&\frac{1}{32(i\omega-\tilde{u}-1)^2}\bigg(\frac{(\tilde{u}+1)(\tilde{u}-3)}{(\tilde{u}-1)^3}+\frac{5}{(2-\tilde{u})(\tilde{u}-1)}+\frac{2\tilde{u}-1}{\tilde{u}(2-\tilde{u})}+\frac{-2\tilde{u}^2+\tilde{u}-1}{\tilde{u}(\tilde{u}-1)^2}\bigg)\nn\\
&+&\frac{\tilde{u}}{(i\omega+\tilde{u}-1)^2}\bigg( \frac{\tilde{u}+(\tilde{u}-1)^2}{32(\tilde{u}-1)^2(2-\tilde{u})}+\frac{-\tilde{u}^3-3\tilde{u}^2+\tilde{u}+1}{8\tilde{u}(\tilde{u}^2-1)^2}+\frac{\tilde{u}-2}{8(\tilde{u}-1)^3}\bigg)\Bigg],\nn\\
\eea

\bea
\Gamma^{(B)}_{4(b)}&=&\frac{(1.15)^2\tilde{t}^4}{\Delta}\Bigg[\frac{1}{2(i\tilde{\omega}-\tilde{u}+1)^2}\bigg(\frac{1}{(i\tilde{\omega}-\tilde{u}+5)(i\tilde{\omega}+\tilde{u}+3)^2}+\frac{1}{4(i\tilde{\omega}-\tilde{u}+5)}+\frac{1}{(i\tilde{\omega}+\tilde{u}+3)(i\tilde{\omega}-\tilde{u}+5)}\nn\\
&+&\frac{(\tilde{u}-1)^3-2\tilde{u}^3}{4(\tilde{u}-1)^2((i\tilde{\omega}+3)^2-\tilde{u}^2)}+\frac{1}{2(\tilde{u}-1)(i\tilde{\omega}+\tilde{u}+3)^2}-\frac{i\tilde{\omega}+2}{4((i\tilde{\omega}+3)^2-\tilde{u}^2)}-\frac{1}{2(\tilde{u}-2)}+\frac{3}{8(\tilde{u}-1)^2}+\frac{1+\tilde{u}}{16\tilde{u}}\bigg)\nn\\
&+&\frac{1}{(i\tilde{\omega}-\tilde{u}+1)}\bigg( \frac{-1}{(i\tilde{\omega}+\tilde{u}+1)(i\tilde{\omega}-\tilde{u}+5)(i\tilde{\omega}+\tilde{u}+3)^2}+\frac{1}{2(\tilde{u}-2)(i\tilde{\omega}+\tilde{u}+1)}-\frac{1}{4(i\tilde{\omega}+\tilde{u}+1)(i\tilde{\omega}-\tilde{u}+5)}\nn\\
&-&\frac{1}{(i\tilde{\omega}+\tilde{u}+1)(i\tilde{\omega}+\tilde{u}+3)(i\tilde{\omega}-\tilde{u}+5)}+\frac{\tilde{u}-3}{16(\tilde{u}-1)(i\tilde{\omega}+\tilde{u}+1)}-\frac{1}{2(\tilde{u}-1)(i\tilde{\omega}+\tilde{u}+1)(i\tilde{\omega}+\tilde{u}+3)^2}\nn\\
&+&\frac{i\tilde{\omega}+2}{4(i\tilde{\omega}+\tilde{u}+1)((i\tilde{\omega}+3)^2-\tilde{u}^2)}+\frac{2\tilde{u}^3+(1-\tilde{u})^3}{4(\tilde{u}-1)^2(i\tilde{\omega}+\tilde{u}+1)((i\tilde{\omega}+3)^2-\tilde{u}^2)}-\frac{2\tilde{u}^3+(1-\tilde{u})^3}{16(\tilde{u}-1)^3(i\tilde{\omega}+\tilde{u}+1)}\nn\\
&&\qquad\qquad+\frac{\tilde{u}-3}{4(\tilde{u}-2)^2}+\frac{7\tilde{u}-5}{64(\tilde{u}-1)^2}-\frac{\tilde{u}(\tilde{u}+2)}{32(\tilde{u}-1)^4}\bigg)\nn\\
&+&\frac{1}{2(i\tilde{\omega}+\tilde{u}+1)^2}\bigg( \frac{1}{(i\tilde{\omega}-\tilde{u}+5)(i\tilde{\omega}+\tilde{u}+3)}+\frac{1}{4(i\tilde{\omega}-\tilde{u}+5)}+\frac{2\tilde{u}^3-(\tilde{u}-1)^3}{4(\tilde{u}-1)^2((i\tilde{\omega}+3)^2-\tilde{u}^2)}-\frac{1}{4(i\tilde{\omega}+\tilde{u}+3)^2}\nn\\
&&\qquad\qquad\frac{1}{2(\tilde{u}-1)(i\tilde{\omega}+\tilde{u}+3)^2}-\frac{(i\tilde{\omega}+2)}{4((i\tilde{\omega}+3)^2-\tilde{u}^2)}-\frac{1}{2(\tilde{u}-2)}+\frac{2-\tilde{u}}{8(\tilde{u}-1)}+\frac{\tilde{u}^3}{4(\tilde{u}-1)^3}\bigg)\nn\\
&+&\frac{1}{(i\tilde{\omega}+\tilde{u}+1)}\bigg(\frac{\tilde{u}-3}{32(\tilde{u}-1)}+\frac{\tilde{u}-3}{8(\tilde{u}-2)^2}-\frac{\tilde{u}^3(\tilde{u}-2)}{32(\tilde{u}-1)^4}\bigg)\Bigg].\nn\\
\eea
where  as before, $\tilde{t}=\frac{t}{\Delta}, \tilde{\omega}=\frac{\omega}{\Delta}$ and $\tilde{u}=\frac{u}{\Delta}$.
\end{widetext}

\end{document}